# Self-Assembled, Nanostructured, Tunable Metamaterials via Spinodal Decomposition

Zuhuang Chen,[*,†,‡] Xi Wang,[†] Yajun Qi,[§] Sui Yang,[‡,∥] Julio A. N. T. Soares,[#] Brent A. Apgar,[†] Ran Gao,[†] Ruijuan Xu,[†] Yeonbae Lee,[†] Xiang Zhang,[‡,∥] Jie Yao,[†,‡] and Lane W. Martin[*,†,‡]

[†] Department of Materials Science and Engineering, University of California, Berkeley, Berkeley, CA 94720, USA

[‡] Materials Science Division, Lawrence Berkeley National Laboratory, Berkeley, CA 94720, USA

[§] Hubei Collaborative Innovation Centre for Advanced Organic Chemical Materials, Key Laboratory of Green Preparation and Application for Materials, Ministry of Education, Department of Materials Science and Engineering, Hubei University, Wuhan 430062, P. R. China

[∥] NSF Nanoscale Science and Engineering Center (NSEC), University of California, Berkeley, Berkeley, CA 94720, USA

[#] Materials Research Laboratory, University of Illinois, Urbana-Champaign, Urbana, IL 61801, USA

**Corresponding Author:**

* E-mail: zuhuang@berkeley.edu, lwmartin@berkeley.edu

**Abstract**

Self-assembly *via* nanoscale phase-separation offers an elegant route to fabricate nanocomposites with physical properties unattainable in single-component systems. One important class of nanocomposites are optical metamaterials which exhibit exotic properties and lead to opportunities for agile control of light propagation. Such metamaterials are typically fabricated *via* expensive and hard-to-scale top-down processes requiring precise integration of dissimilar materials. In turn, there is a need for alternative, more efficient routes to fabricate large-scale metamaterials for practical applications with deep-subwavelength resolution. Here, we demonstrate a bottom-up approach to fabricate scalable nanostructured metamaterials *via* spinodal decomposition. To demonstrate the potential of such an approach, we leverage the innate spinodal decomposition of the $VO_2$-$TiO_2$ system, the metal-to-insulator transition in $VO_2$, and thin-film epitaxy, to produce self-organized nanostructures with coherent interfaces and a structural unit cell down to 15 nm (tunable between horizontally- and vertically-aligned lamellae) wherein the iso-frequency surface is temperature-tunable from elliptic- to hyperbolic-dispersion producing metamaterial behavior. These results provide an efficient route for the fabrication of nanostructured metamaterials and other nanocomposites for desired functionalities.

Next-generation opto-electronic applications require diverse and nanoscale materials in complex geometries to achieve desired functionalities. For example, metamaterials, which are artificial composites with structural unit cells much smaller than the wavelength of light,[1] are poised to impact such opto-electronic applications. Metamaterials leverage a complex mixture of properties in specific geometries to realize exotic optical properties, including negative refraction,[2] optical magnetism,[3] and hyperbolic dispersion,[4] which could enable a plethora of exciting applications such as subwavelength imaging[5] and invisibility cloaking.[6] Such applications require large-scale or even three-dimensional bulk metamaterials.[7] To achieve deep subwavelength-scale features, multilayer deposition,[8,9] nano-imprinting,[10] and layer-by-layer lithography techniques[11] have been explored to fabricate three-dimensional metamaterials. Roughness induced at each interface in these techniques is, however, accumulated thus exacerbating scattering loss.[12] To date, most metamaterials are based on metal-dielectric composite structures which, in turn, presents considerable materials (*e.g.*, typical metals such as silver that are used have limited tunability of their optical responses) and fabrication (*e.g.*, production of continuous, high-quality sub-20 nm thick metal films in intimate contact with ceramic or semiconducting dielectric layers is difficult) challenges. As a result, an open challenge is the production of tunable metamaterials operating in the visible and near-infrared wavelength regimes since this requires the fabrication of nanostructures with sharp interfaces and periodicities on the order of a few tens of nanometers over macroscopic areas.[7,12]

At the same time, the study of nanoscale phase-separation has advanced rapidly because the resulting self-organized nanocomposites can exhibit a variety of coupled functionalities.[13] In recent years, considerable effort has focused on self-assembled oxide nanocomposites and their potential applications in nanoelectronic devices.[14-17] Similar bottom-up, self-assembly techniques, such as electrochemically grown metal wire arrays,[18] block copolymer self-assembly,[19] DNA-mediated self-assembly of nanoparticles,[20] feedback-driven self-assembly,[21] and one-step pulsed-laser deposition,[22] have been explored for the fabrication of metamaterials. Most of these self-assembly approaches are, however, either template-assisted methods, which are complicated and require additional post-possessing steps, such as etching and metal deposition or solution-based approaches which are not compatible with semiconductor integration procedures. More recently, self-assembly of eutectic structures has been explored to fabricate metamaterials, but the resulting length scales are > 200 nm and all the components are limited to dielectric materials; thus limiting their utility.[23,24] Self-assembly *via* spinodal decomposition, however, could offer an innovative route to scalable fabrication of metamaterials as well as controlled structural features at the nanoscale.[13,25] Spinodal decomposition is a phase-separation process whereby a material spontaneously separates into two phases with distinct compositions.[26,27] Unlike conventional phase separation which occurs *via* nucleation and growth, spinodal decomposition does not require nucleation and is solely determined by diffusion; leading to the spontaneous formation of structures with compositional fluctuations on the nanometer length scale.[28,29] Furthermore, since spinodal decomposition is a continuous process, the interfaces between the two separated phases remain coherent,[30] which could effectively reduce light scattering in optical structures. Spinodal decomposition is a ubiquitous phenomenon occurring in a diverse set of systems including metal alloys,[27] oxides,[28,31] semiconductors,[32] and polymers[33] which could provide for a range of material choices for optical applications. All told, these characteristics provide a number of potential advantages for the fabrication of self-assembled nanocomposite metamaterials and other multifunctional applications.

Here, we demonstrate a simple bottom-up approach to create self-assembled, nanostructured metamaterials with controllable structural geometry (*i.e.*, horizontally- or vertically-aligned lamellae) and temperature-tunable optical response from spinodally-decomposed $VO_2$-$TiO_2$ epitaxial thin films. The $VO_2$-$TiO_2$ system was explored for several reasons: 1) the system is known to exhibit a spinodal instability,[34] 2) the component materials have vastly different optical properties making it potentially interesting from a metamaterial standpoint, and 3) the well-known metal-to-insulator transition in $VO_2$ has been widely studied and gives rise to optical tunablity.[35-37] To date, what little work exists on $VO_2$-based metamaterials has focused on single-layer $VO_2$ as a tunable-substrate for patterned antennas and, as a result of the relatively weak coupling/interaction, has demonstrated limited utility and tunability.[38,39] In our work, as-grown solid-solution films are driven to phase separate upon post-annealing and we demonstrate the ability to deterministically create horizontally- or vertically-aligned lamellae consisting of Ti- and V-rich phases depending on the substrate orientation wherein the composition modulation is always along the elastically-soft $[001]_S$ (where S refers to the substrate reference frame). These lamellae have coherent interfaces and characteristic length scales as small as ~15 nm; smaller than what can be achieved *via* conventional top-down methods. In turn, by taking advantage of the metal-to-insulator phase transition that occurs in the V-rich phase (similar to that in $VO_2$) just above room temperature, the optical iso-frequency surface of the self-assembled nanostructures can be made to exhibit a temperature-tunable transition from elliptic to hyperbolic dispersion in the near-infrared range and thus the formation of hyperbolic metamaterial response.

**Results and Discussion**

To begin, the pseudo-binary phase diagram for the $VO_2$-$TiO_2$ system reveals complete solid solubility and a tetragonal, rutile structure at high temperatures, but upon transitioning below ~830 K, a spinodal instability gives rise to a miscibility gap centered about a composition of ~35 mol% Ti (Figure 1a).[34] Spinodal instabilities are not rare in materials, but most systems consist of two phases with similar crystal structure and physical properties;[28,29,31,40] thus making the functional properties of many spinodally-decomposed systems rather mundane. In $VO_2$-$TiO_2$, however, the $VO_2$ is a correlated electron system that undergoes a phase transition near room temperature from a high-temperature metallic tetragonal (rutile) phase to a low-temperature insulating monoclinic phase.[35] Associated with this transition is a dramatic change in the electronic conductivity as evidenced by the widely studied metal-to-insulator transition (Figure 1b). At the same time, the optical properties of $VO_2$ also change dramatically. The optical dielectric constants of $VO_2$ films were probed at 303 K and 363 K *via* spectroscopic ellipsometry and the results were fit to extract the optical dielectric constant (real ($\varepsilon'$) and imaginary ($\varepsilon''$) components; details of the measurement and fitting are provided, Experimental Section). These studies show that the room-temperature, insulating, and monoclinic phase of $VO_2$ has a positive dielectric constant (from 600-1600 nm), but the high-temperature, metallic, tetragonal phase exhibits a negative dielectric constant for wavelengths $\gtrsim$ 1000 nm (Figure 1c). Note that for brevity, we show here only $\varepsilon'$, but $\varepsilon''$ was also measured (Supporting Information, Figure S1a). The dramatic differences in resistivity and optical constant between the insulating and metallic phases of $VO_2$ make it a potential candidate for use in tunable plasmonic materials.[36-38] On the other hand, $TiO_2$, which also exhibits a tetragonal rutile structure near room temperature, is a wide band gap, dielectric which exhibits (essentially) temperature-independent positive $\varepsilon'$ across the same temperature and wavelength regimes (Figure 1c and Supporting Information, Figure S1b).

To probe the potential for spinodal decomposition-driven nanostructure formation in this system, we grew films from a $V_{0.6}Ti_{0.4}O_2$ target on rutile $TiO_2$ (001) and (100) single crystal substrates by pulsed-laser deposition at 673 K (see Experimental Section for details). Detailed structural analysis *via* X-ray diffraction and transmission electron microscopy (TEM) reveal that the as-grown $V_{0.6}Ti_{0.4}O_2/TiO_2$ (001) and (100) heterostructures are fully-epitaxial and single-phase solid solutions (Supporting Information, Figure S2). The as-grown heterostructures are likely stabilized as solid solutions due to kinetic limitations in the non-equilibrium growth process (*i.e.*, similar to quenching in bulk samples).[34] This homogeneous solid solution is, however, metastable and will decompose into phase-separated structures given sufficient activation of the diffusion process. In the current work, *ex post facto* annealing at 673 K was used to explore the phase evolution (Figure 2a, b). Upon commencing the anneal, the as-grown, single-phase $V_{0.6}Ti_{0.4}O_2/TiO_2$ (001) heterostructures (Figure 2a) quickly start the decomposition process and after ~5 hours the diffraction peak arising from the solid-solution has become weaker, broader, and has shifted to higher $2\theta$ values (indicating a reduction in the out-of-plane lattice parameter). At the same time, two broad diffraction peaks, on either side of the solid-solution peak, appear as an indication of the onset of phase separation. The low- and high-angle diffraction peaks correspond to Ti- and V-rich phases, respectively. Further annealing (*e.g.*, 24 hours) results in a nearly complete loss of the solid-solution peak and further growth of the satellite peaks; consistent with what is expected for spinodal decomposition.[40] On the other hand, even after annealing at 673 K for 24 hours, the $V_{0.6}Ti_{0.4}O_2/TiO_2$ (100) heterostructures (Figure 2b) are found to exhibit only a single (slightly shifted) sharp peak. These results suggest that the structural modulation in this case, if it has occurred, does not align along the out-of-plane direction. Reciprocal space mapping (RSM) studies about the 301-diffraction conditions of the film and substrate reveal the presence of satellite peaks in the in-plane direction ($[001]_S$) which indicate a structural modulation with an average periodicity of ~17 nm along the $[001]_S$ (inset, Figure 2b and Supporting Information, Figure S3). Further evidence of the compositional and structural evolution of these heterostructures is obtained from electronic transport measurements. From the phase-diagram (Figure 1a), at room temperature the Ti-rich phase should correspond to $V_{0.35}Ti_{0.65}O_2$ and should exhibit insulating, dielectric behavior while the V-rich phase corresponds to $V_{0.89}Ti_{0.11}O_2$ and should exhibit a metal-to-insulator transition similar to that in $VO_2$ wherein it behaves as a metal at high temperature and as an insulating dielectric at room temperature.[34,41] Subsequent measurements reveal that the as-grown, solid-solution heterostructures exhibit semiconducting transport while the phase-separated heterostructures exhibit sharp metal-to-insulator transitions (Figure 2c). This indicates that the spinodal decomposition process creates a V-rich phase which exhibits the expected metal-to-insulator transition albeit reduced in magnitude compared to the pure $VO_2$ because of the Ti incorporation. This is consistent with prior studies[42,43] which have shown that alloying Ti into $VO_2$ gives rise 1) higher transition temperatures, 2) smaller magnitude changes in resistance across the metal-to-insulator transition, 3) less hysteresis for the metal-to-insulator transition and has been attributed to the added Ti distorting the $VO_6$ octahedra in the parent rutile structure. Such a reduced hysteresis width could also increase the sensitivity of optical responses to temperature.

To further understand the structural evolution, bright-field transmission electron microscopy (TEM) and high-angle annular dark-field scanning transmission electron microscopy (HAADF-STEM) studies were completed on the post-annealed, phase-separated films. Although a time evolution of the phase separation process is provided (Supporting Information, Figure S4), here we focus on phase-separated films after 24 hours of annealing to illustrate the final nanostructures that are obtained. Low-magnification, cross-sectional imaging of the annealed (001)-oriented heterostructures reveals farily-well-ordered, horizontally-layered nanostructures with periodicities of 15-20 nm along the out-of-plane $[001]_S$ (Figure 3a). Low-magnification, cross-sectional imaging of the annealed (100)-oriented heterostructures confirms the presence of vertically-aligned nanostructures with periodicities of 15-20 nm along the in-plane $[001]_S$ (Figure 3b). Plan-view imaging of the same (100)-oriented heterostructures further reveal the long axes of the lamellae running along the $[010]_S$ (Supporting Information, Figure S5). Subsequent STEM-based energy-dispersive X-ray spectroscopy studies reveal that the lamellae consist of V- (blue regions, Figure 3c, d) and Ti-rich (green regions, Figure 3c, d) phases with composition modulation along the $[001]_S$. High-resolution imaging near interfaces between the V- and Ti-rich phases (Figure 3e, f) confirms the presence of (001)-type interfaces and that the boundaries between the two phase-separated constituents are coherent – meaning that despite a rather large difference in lattice parameter, the interfaces remain pristine at this atomic level with no obvious dislocations or extended defects. Examination of fast-Fourier transform (FFT) patterns of the Ti-rich regions (top inset, Figure 3e) reveals that the Ti-rich phase exhibits tetragonal structure while the V-rich phase (bottom inset, Figure 3e) exhibits extra superlattice spots indicative of the presence of a monoclinic distortion.[41] The monoclinic nature of the V-rich phase was further confirmed by selected area electron diffraction (SAED) (inset, Figure 3b). The SAED pattern shows extra superlattice spots in addition to the fundamental Bragg reflections which are attributed to the monoclinic structure.[44] No superlattice spots are observed in the SAED pattern of as-grown, solid-solution heterostructures (Supporting Information, Figure S2h).

To summarize, at this point we have demonstrated the ability to produce self-assembled, quasi-periodic nanocomposite thin films with coherent interfaces, controlled structural motifs (*i.e.*, horizontally- and vertically-aligned lamellae), and unit-cell dimension of 15-20 nm. The unidirectional structural/compositional modulation occurring along the [001]$_S$ is likely due to the anisotropic elastic (strain) energy that works to maintain coherency between the two resultant phases (note that the misfit strain along the [100]$_S$ or [010]$_S$ is smaller than that along the [001]$_S$ axis).[41] Based on this interesting structural and, in turn, electronic phase separation we have gone on to probe the optical properties of these phase-separated nanostructures.

To investigate the potential of these nanostructures for metamaterial behavior, the optical dielectric constants were probed using spectroscopic ellipsometry (see Experimental Section).[45,46] For the as-grown, homogenous, solid-solution films, fittings were completed assuming isotropic response (*i.e.*, $\varepsilon_{[100]} = \varepsilon_{[010]} = \varepsilon_{[001]}$ where $\varepsilon_{[uvw]}$ is the optical dielectric constant along the [uvw] and is equal to $\varepsilon' + i\varepsilon''$). For simplicity, we provide data for $\varepsilon'$ only in the main text, but data for $\varepsilon''$ is also available (Supporting Information, Figure S6). The resulting optical dielectric constant for the as-grown, solid-solution was found to be positive throughout the visible and near-infrared wavelength regimes at all temperatures (blue-dashed line, Figure 4a-d); consistent with the observed semiconducting behavior. For the phase-separated nanostructures, however, one can no longer assume that the heterostructures are isotropic. Thus for the (100)- and (001)-oriented heterostructures, fittings were done assuming uniaxial anisotropy wherein $\varepsilon_{[100]} = \varepsilon_{[010]} \neq \varepsilon_{[001]}$ since the structural and compositional modulation is always along the [001]$_S$. Measurements at 303 K, where both the V- and Ti-rich phases are non-metallic, reveal that $\varepsilon'_{[100]}$, $\varepsilon'_{[010]}$, and $\varepsilon'_{[001]}$ are all positive for both the (100)- and (001)-oriented heterostructures (Figure 4a and b, respectively). Measurements at 363 K, wherein the V-rich phase is now metallic, however, show something very different. For both the (100)- and (001)-oriented heterostructures, wherein the lamellae are vertically- and horizontally-aligned relative to the film surface, respectively, the $\varepsilon'_{[100],[010]}$ and $\varepsilon'_{[001]}$ responses show counter-trending behavior with $\varepsilon'_{[001]}$ remaining positive across the entire wavelength regime studied herein and $\varepsilon'_{[100],[010]}$ trending towards negative and eventually turning negative at wavelengths >1480 nm in the (001)-oriented heterostructures (Figure 4c,d). In other words, the combination of the spinodal-decomposition which drives nanostructuring and the temperature-driven metal-to-insulator transition in the V-rich phase, creates a mesoscale system which is fundamentally different at high and low temperatures and from the as-grown, solid solution. The presence of nanoscale order of the metallic and dielectric phases, with a unit cell periodicity much shorter than the wavelength of light, drives metamaterial response including further enhancement of the uniaxial anisotropy of the optical response. In fact, one can reproduce similar trends as the ellipsometrically-derived data using a simple effective medium approximation considering a layered $VO_2$-$TiO_2$ structure (since the periodicity is much shorter than the wavelength of light) (Supporting Information, Figure S7). The effective medium model, in turn, lends credence to the observed trends in anisotropic dielectric response in our self-assembled nanostructures. Finally, we note that although at 363 K $\varepsilon_{[100]}$ and $\varepsilon_{[010]}$ for the (100)-oriented heterostructures show a trend toward negative values at longer wavelengths (Fig. 4c) where hyperbolic dispersion is promised, the failure to achieve actual negative values in this wavelength regime can likely be ascribed to the less ordered nature of these structures. Suggesting that future work to improve structural order could further improve the optical response of these structures.

To further understand the nature of this enhanced optical anisotropy, we explored the optical iso-frequency surface of the various materials. In general, the electromagnetic wave ($k$-wavevector) propagation through uniaxially-anisotropic materials where $\varepsilon_{[100]} = \varepsilon_{[010]} \neq \varepsilon_{[001]}$ is governed by the dispersion relation:

$$\frac{k_{[100]}^2 + k_{[010]}^2}{\varepsilon_{[001]}} + \frac{k_{[001]}^2}{\varepsilon_{[100],[010]}} = \frac{\omega^2}{c^2} \tag{1}$$

where $k_{[uvw]}$ is the [uvw] component of the wave vector, $\omega$ is the frequency, and $c$ is the speed of light. Using optical properties at a wavelength of 1700 nm as an example, at 303 K, the as-grown, solid solution (which is isotropic in nature) has an iso-frequency $k$-space surface which is spherical while the phase-separated heterostructures (which are anisotropic in nature) exhibit an ellipsoidal iso-frequency $k$-space surface (Figure 4e). At 363 K, in the phase-separated, (001)-oriented heterostructures, however, we observe an extreme in dielectric anisotropy where in $\varepsilon'_{[100],[010]}$ and $\varepsilon'_{[001]}$ have opposite signs. In such a case, the iso-frequency surface will become an open hyperboloidal shape.[47] In turn, the self-assembled, horizontally-aligned nanocomposite is thus characterized as a tunable metamaterial which exhibits a dramatic change in the wavevector iso-frequency contour from a closed ellipsoid at low temperature (Figure 4e) into an open hyperboloid at high temperature (Figure 4f) wherein negative refraction could be possible (Supporting Information, Figure S8).[48] Such hyperbolic dispersion could support infinitely large wavevectors (known as high-$k$ waves) and an enhanced photonic density of states as compared to conventional materials which have a bounded spherical/elliptical iso-frequency surface.[49] In turn, such effects have garnered much of the attention given to hyperbolic metamaterials which are considered potentially exciting for a range of optical imaging, sensing, and emission engineering applications.[8,18,47,48] For example, by supporting the propogation of evanscent waves, hyperlenses based on hyperbolic metamaterials can magnify sub-diffraction-limited objects (*i.e.*, far-field sub-wavelength imaging).[50]

**Conclusion**

All told, such self-assembled, ordered-nanocomposite films with tunable structural motifs and unit-cell dimensions as small as ~15 nm (smaller than the critical thickness for growth of uniform-continuous metallic film on semiconductor/dielectric substrates and that achieved by conventional top-down process) can be obtained by leveraging the innate tendencies of materials to phase separate as in the spinodal decomposition of the $VO_2$-$TiO_2$ system studied herein. Through the use of thin-film epitaxy, we gain additional control of the phase separation whereby unidirectional decomposition along the $[001]_S$ is observed and thus by changing substrate orientation, we can create horizontally- or vertically-aligned lamellae. Subsequent optical characterization of these phase-separated materials demonstrates the potential of this approach to enhance optical dielectric anisotropy and even gives rise to reversible, temperature-tunable transformation from non-hyperbolic to hyperbolic metamaterial behavior in the near-infrared wavelength regime. The ability to tune the iso-frequency surface from non-hyperbolic to hyperbolic provides additional degrees of freedom to control light-matter interaction. In addition to the global thermal triggering used here, electrical heating,[51] electric field,[52,53] and optical excitation,[54,55] could provide alternative ways to induce phase transition and control the iso-frequency surface in the self-assembled nanostructured metamaterials.

This approach offers an alternative to complex and expensive top-down fabrication methods for the production of nanostructured metamaterials. With future efforts more ordered structures could be achieved and alternative material systems (such as Al-Si[56] and $TiO_2$-$RuO_2$[57]) and other geometries (such as 1-3-type nanowire structures) could be identified which would further enhance the efficacy of this spinodal-decomposition process for the scalable production of metamaterials. In addition, it is conceivable that one could greatly increase the scale of these structures while keeping the same feature size or even synthesize bulk versions of these materials to produce three-dimensional metamaterials. In summary, this work provides a approach by which one can fabricate large-scale, three-dimensional metamaterials with nanoscale features and addresses challenges in the community associated with making these exciting materials. Finally, these results also have broad implications for the fabrication of controlled, self-organized nanocomposite thin films for desired functionalities using spinodal instabilities.

**Methods**

**Thin-film growth.** $VO_2$ and $V_{0.6}Ti_{0.4}O_2$ films with thickness of 50-90 nm were grown on rutile $TiO_2$ (100) and (001) single-crystal substrates by pulsed-laser deposition from ceramic targets of the same composition at 673 K, an oxygen pressure of 10 mTorr, at a laser repetition rate of 5 Hz, and a laser fluence of ~1.2 J/cm$^2$. Following growth, the heterostructures were cooled to room temperature at a rate of 5°C/min. under a dynamic oxygen pressure of 10 mTorr. *Ex post facto* annealing was also completed at 673 K and in an oxygen pressure of 10 mTorr for up to 24 hours.

**Structural and transport characterization.** X-ray $\theta - 2\theta$ and reciprocal space mapping studies were carried out on an X'Pert MRD Pro diffractometer (Panalytical) using momchromatic Cu-$K_\alpha$ radiation. Thicknesses were measured *via* fitting of X-ray diffraction Laue fringes and X-ray reflectivity studies. Cross-sectional TEM samples were prepared using standard procedures including cutting, gluing, mechanical polishing, and ion milling. The ion milling process was performed on a Precision Ion Polishing System (PIPS, model 691, Gatan) with an incident ion angle of 5° and an accelerating voltage of 3 kV using liquid $N_2$ to cool the stage. Bright-field TEM investigations were carried out on a Tecnai G20 and HAADF-STEM was completed in a FEI TITAN Cs-corrected ChemiSTEM electron microscope operated at 200 kV. The temperature-dependent transport studies were completed in a van der Pauw configuration in a Quantum Design Physical Property Measurement System.

**Optical characterization.** Variable angle ellipsometry measurements were completed in a variable angle spectroscopic ellipsometer (VASE, J. A. Woollam Co, Inc.) with a heater cell in the spectral range from 400-1700 nm. The incident angle was varied from 45°-75° with a step size of 5°. The ellipsometry measurements provide two parameters $\varphi$ and $\Delta$. These ellipsometry parameters are, in turn, related to the ratio of the reflection coefficients for the light of p-polarization $r_p$ and s-polarization $r_s$: $\rho(\theta_i) = \frac{r_p}{r_i} = \tan(\varphi)\exp(i\Delta)$. In turn, the optical dielectric constants were obtained by fitting the ellipsometry data using different models to generate best-fits to the measured $\varphi$ and $\Delta$ using the VASE software (raw ellipsometry data of the parameter φ vs. wavelength is provided, Supporting Information, Figure S9, S10, and Figure S11). The insulating $TiO_2$ was fitted by a biaxial layer with two Cauchy models, since the rutile $TiO_2$ has anisotropic dielectric response (*i.e.*, $\varepsilon_{[100]} = \varepsilon_{[010]} \neq \varepsilon_{[001]}$).[58] A General Oscillator layer (GOL) model[59] consisting of two Lorentz oscillators (Lorentz model) was used to fit the insulating $VO_2$ at 303 K and the as-grown, solid-solution heterostructures (semiconducting in nature) at both temperatures. A GOL model consisting of one Lorentz oscillator and one Drude oscillator (Drude-Lorentz model) was used to fit the metallic $VO_2$ at 363 K. A biaxial layer with two GOLs consisting of hybrid Gaussian-Lorentz-Drude oscillators was used to fit all phase separated structures.[42,43] To simplify the complexity of fitting, we started our fitting from 600 nm to reduce the number of oscillators in short wavelength region. All fitting resulted in a reasonably small mean squared error (less than 3).

ASSOCIATED CONTENT

**Supporting Information.** The supporting information section provides information of optical dielectric response of $TiO_2$ and $VO_2$ thin films, structural characterization of as-grown $V_{0.6}Ti_{0.4}O_2$ thin films, reciprocal space mapping of phase-separated (100)-oriented heterostructures, evolution of phase separation during spinodal decomposition, plan-view TEM imaging of the phase-separated (100)-oriented heterostructure, optical dielectric response of various heterostructures, effective medium theory calculations multilayer hyperbolic metamaterials, tunable refraction simulation, and ellipsometric measurements of heterostructures.

This material is available free of charge *via* the Internet at http://pubs.acs.org.

**AUTHOR INFORMATION**

**Corresponding Author:**

Email: zuhuang@berkeley.edu

Email: lwmartin@berkeley.edu

**Notes**

The authors declare no competing financial interest.

**Acknowledgements**

X. W. and Y. Q. contributed equally to this work. We would like to thank Dr. Jingbo Sun for useful discussion. Z.H.C. and R.G. acknowledge the support of the Air Force Office of Scientific Research under grant FA9550-12-1-0471. Y.Q. acknowledges support of the National Science Foundation of China under grant 11204069 and 51472078. B.A.A. acknowledges support from the Department of Energy, Basic Energy Science under grant No. DE-SC0012375 for the development of various oxide films and optical studies. R.X. acknowledges support from the National Science Foundation under grant DMR-1451219. L.W.M. acknowledges support from the Laboratory Directed Research and Development Program of Lawrence Berkeley National Laboratory under U.S. Department of Energy Contract No. DE-AC02-05CH11231 for the development of light-matter interactions in materials. X.W. and J.Y. acknowledge the support from Hellman Family Foundation. The ellipsometry measurements were carried out in the Frederick Seitz Materials Research Laboratory Central Research Facilities, University of Illinois.

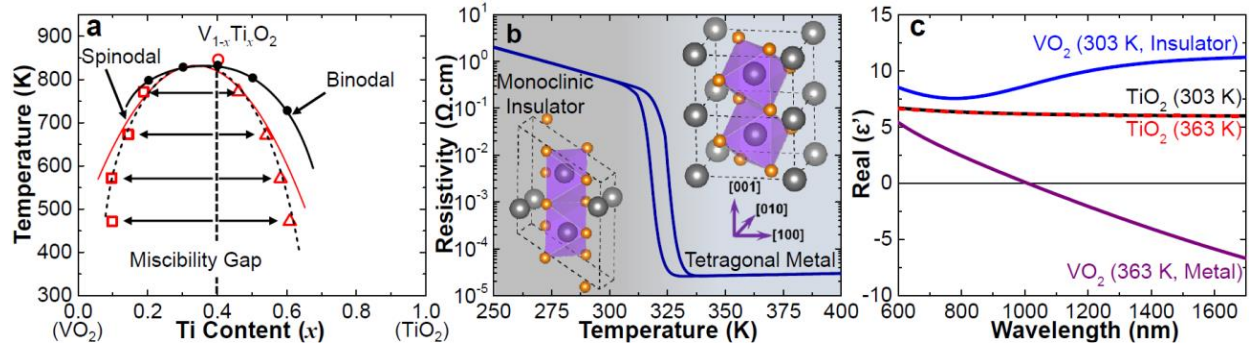

**Figure 1.** VO$_2$-TiO$_2$ spinodal system. (a) Phase diagram of the VO$_2$-TiO$_2$ system (adapted from Ref. [34]). (b) Temperature dependence of the resistivity of a ~70 nm VO$_2$/TiO$_2$ (001) heterostructure wherein the VO$_2$ exhibits a sharp metal-to-insulator transition with a four orders of magnitude change in resistivity at a transition temperature of ~325 K. Insets show the monoclinic (left) and rutile tetragonal (right) unit cells. (c) Real part of optical dielectric constant ($\varepsilon'$) of VO$_2$ and TiO$_2$ at 303 K and 363 K. Of note is the fact that $\varepsilon'$ for VO$_2$ in the metallic state (363 K) is negative at wavelengths ≳ 1000 nm.

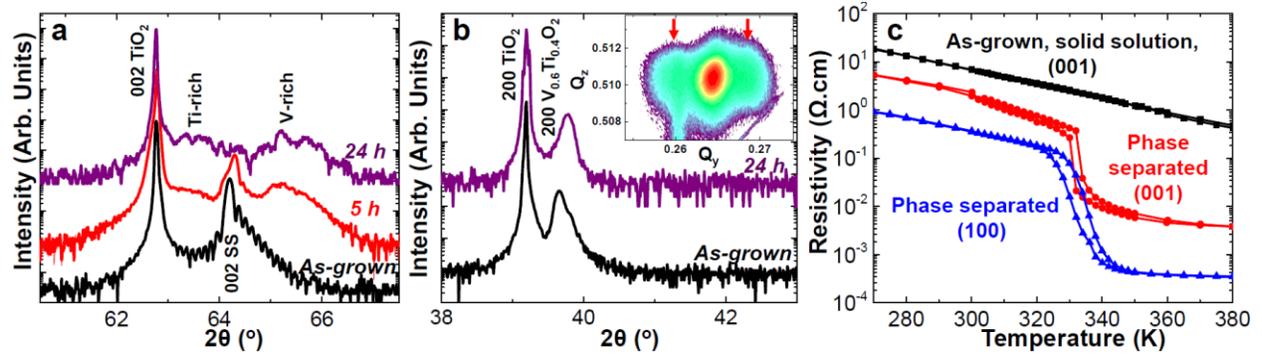

**Figure 2.** Evolution of the spinodal nanostructures. X-ray diffraction of the evolution of the (a) (001)- and (b) (100)-oriented heterostructures with annealing time at 673 K. The inset of (b) shows X-ray reciprocal spacing mapping of the annealed (100)-oriented heterostructure about the 301-diffraction condition wherein satellite peaks with periodicity of ~17 nm are observed along the $[001]_S$. (c) Temperature dependence of the resistivity of the as-grown and spinodally-decomposed films revealing that upon phase separation, the heterostructures form a V-rich phase with a strong metal-to-insulator transition.

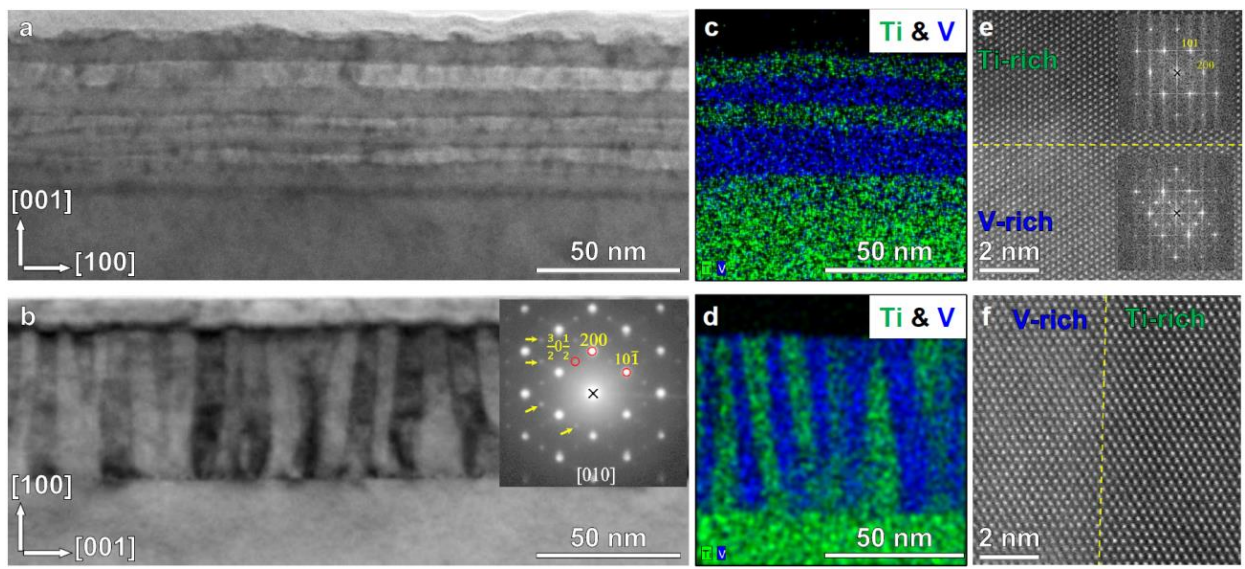

**Figure 3.** Nano- and atomic-scale structure of phase-separated heterostructures. Cross-sectional bright-field TEM images of the (a) (001)- and (b) (100)-oriented heterostructures after annealing at 673 K for 24 hours. The inset of (b) shows a SAED pattern taken from the top area of the film wherein the yellow arrows denote the superlattice peaks arising from the monoclinic nature of the V-rich phase. Combined STEM image and EDS elemental maps where in V is marked as blue and Ti as green for the annealed (c) (001)- and (d) (100)-oriented heterostructures. HAADF-STEM images of the interface between the V- and Ti-rich phases in the (e) (001)- and (f) (100)-oriented heterostructures. The insets of (e) show the FFT pattern of the Ti-rich area (top) and the V-rich phase area (bottom) from which the Ti-rich phase is found to be tetragonal and the V-rich phase is to be monoclinic in nature.

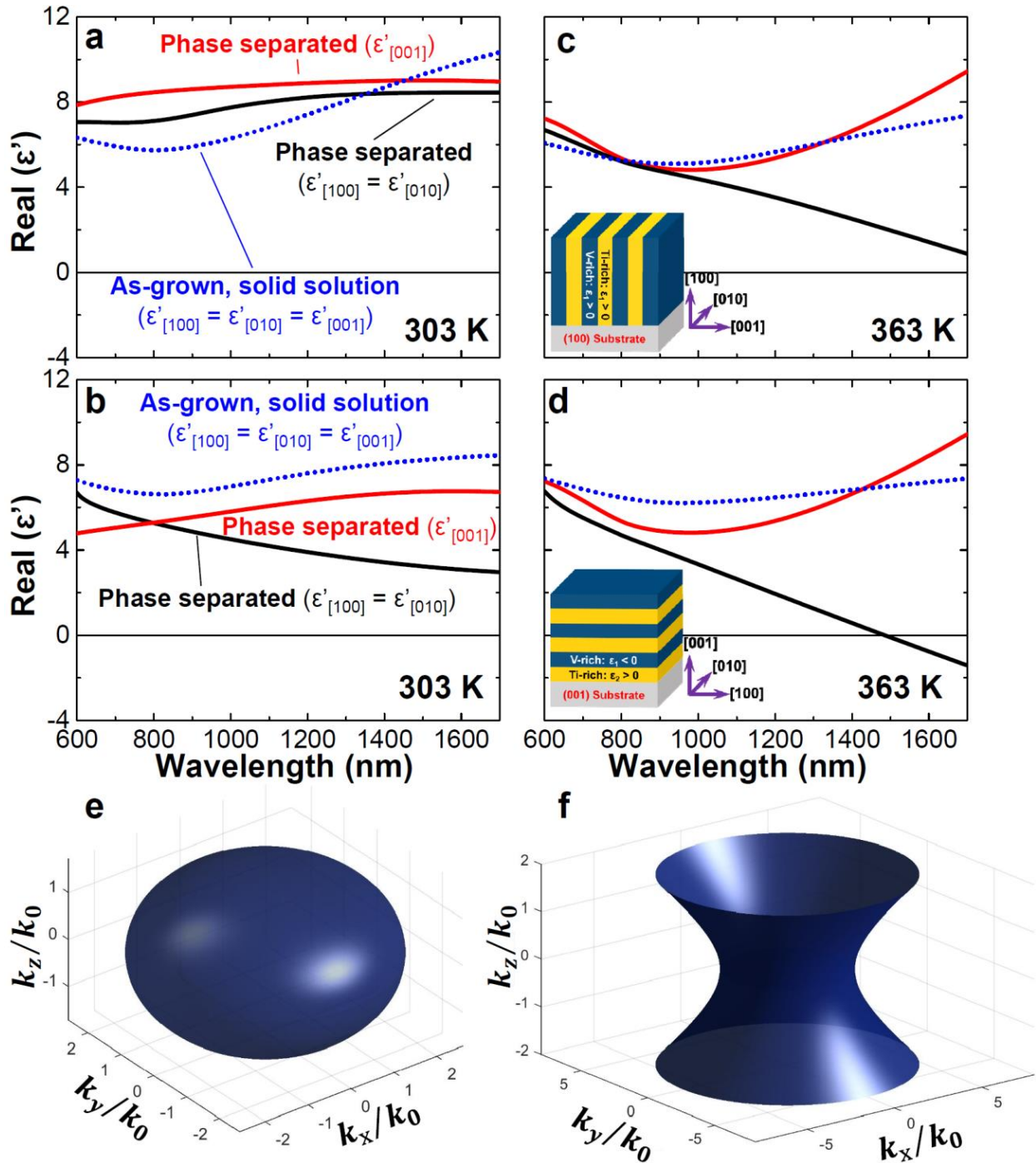

**Figure 4.** Nanostructure optical properties. Various components of the real part of the optical dielectric constant ($\varepsilon'$) measured as a function of wavelength for as-grown, solid-solution (SS) heterostructures (dashed blue) and phase-separated nanostructures (solid black and red) for (a) (100)- and (b) (001)-oriented heterostructures at 303 K and for (c) (100)- and (d) (001)-oriented

heterostructures at 363 K. The extracted iso-frequency surfaces of self-organized horizontally aligned nanostructures (at 1700 nm) at (e) 303 K and (f) 363 K where $k_o = \frac{\omega^2}{c^2}$ is the wavenumber in vacuum, $k_x$, $k_y$ and $k_z$ are the [100], [010] and [001] components of the wave vector, respectively.